\documentclass[12pt,a4]{article}
\usepackage{graphicx}
\usepackage{subfigure}

\input tcilatex
\QQQ{Language}{
American English
}

\begin{document}

\title{{\Large {\bf \ First principles calculations of monolayer compressibilities }%
}}
\author{{\normalsize \ Cristi\'{a}n Sanchez and Ezequiel Leiva\thanks{{Corresponding
author. Fax: +54 51 334174; E-mail: eleiva@fcq.uncor.edu}}} \\
{\normalsize Unidad de Matem\'{a}tica y F\'{\i}sica, Facultad de Ciencias
Qu\'{\i}micas}\\
{\normalsize Universidad Nacional de C\'{o}rdoba. C.C. 61, A.P. 4, 5000
C\'{o}rdoba}\\
{\normalsize \ Argentina }}
\maketitle

\begin{abstract}
We perform high quality, first principles calculations of the properties of
Pb and Tl isolated monolayers. Among these, we consider the equilibrium
lattice constant, the two dimensional compressibilities and the electronic
density. Comparison is made with previous results obtained using more
simplified models. The present results represent an improvement concerning
the calculated compressibilities; these remaining still lower than the
measured values. We speculate that the latter could be due to some
corrugation of the monolayer, not considered in the present modeling.

\newpage\ 
\end{abstract}

\section{ Introduction}

The application of in-situ surface X-ray scattering techniques to the study
of electrochemical systems has contributed to the elucidation of the
structure of metal monolayers adsorbed at underpotential \cite{Toney1,
Toney2, Toney3, Toney4, Toney5, Toney6}. Among these systems is particularly
interesting the case of sp metals adsorbed at underpotential on dense
substrate faces like Ag(111) and Au(111). In these cases, the adlayer
presents an incommensurable structure, whose lattice constant varies with
the applied electrode potential. The monolayer is built with a lattice
constant which is smaller than that expected from the bulk structure. This
contraction of the adsorbed monolayer has been theoretically explained by
Leiva and Schmickler \cite{L&S95} by means of a jellium model with
pseudopotentials, using the density functional formalism to write down the
energy of the system. In the case of the detachment of a single slab from
the bulk, it was found that the electronic density of the isolated layer
expanded in the direction perpendicular to the lattice plane. This
relaxation, concomitant with a decrease of the nearest neighbor distance
between the ion cores in the slab, occurs because the absence of the ions of
neighboring planes sitting on hollow sites reduces the electronic pressure
in the direction perpendicular to the slab. In the case of the experiments,
it was found that when the potential is shifted cathodically, more atoms are
packed into the adsorbed monolayer, with the concomitant decrease of the
lattice constant. From this experimental information, the 2D isothermal
compressibility $K_{2D}$ is estimated to be in the range of 1.2 to 1.9 $\AA
^2/eV.$ A first estimation of this compressibility in terms of a free
electron model yielded values in the order of 0.5 $\AA ^2/eV$, that is,
considerably lower than in the experiment. Recent calculations by Leiva and
Schmickler \cite{L&S97} using a more sophisticated jellium-pseudopotential
model also rendered values which were low as compared with the experiment.
It is the purpose of this work to perform high quality, first principles
calculations for some of the systems addressed above, which should clarify
if the relatively low theoretical compressibilities are due to
simplifications in the model or to some other aspects so far not taken into
account. We consider here the cases of Pb and Tl, which are the systems for
which a wealth of information exists.

\begin{figure}
  \centering
  \includegraphics[scale= .40,angle=90]{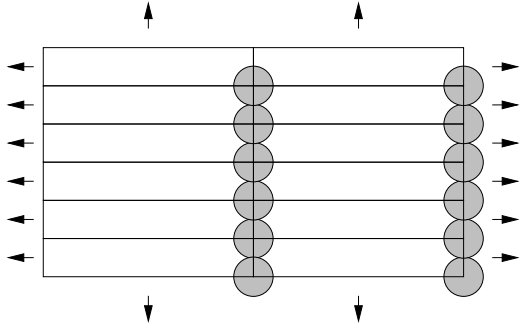}
  \caption{Schematic illustration of the supercell geometry employed
in this work in order to represent a single layer of a metal isolated in
vacuum.}
\end{figure}

\section{Model and calculation method}

In these preliminary calculations we only consider a metal slab isolated in
vacuum, which we contract isotropically, thus obtaining the energy as a
function of the distance between nearest neighbors.

We performed two sets of calculations. One of them was performed by means of
a computer code developed by one of us \cite{Leiva_scf}, which solves an
effective one dimensional Schr\"{o}dinger equation in the presence of
averaged local pseudopotentials \cite{Lang&Kohn} within the density
functional approach \cite{H&K}. Within this model, which will be referred as 
{\it 1-D model} onwards, the electronic density is a function of the
distance perpendicular to the metal surface. The other set of calculations
was performed with the complex computer code fhi96md \cite{fhi}, developed
at the Fritz-Haber-Institut by Scheffler and coworkers. In this case, the
geometry actually corresponds to that of a periodic supercell as
schematically shown in Figure 1 and the pseudo electronic density contains
the full three dimensional features outside a selected core region. We shall
refer to the results obtained by this method as those of the {\it 3-D model}%
.  The geometry of the nuclei is contained in a supercell, which is
periodically repeated as a lattice. Thus, the coordinates of a nucleus or
its periodic image $R_i$ will be

\[
R_i=\tau _i+R 
\]
where $\tau _i$ represents the position of nucleus ``$i$'' inside the
supercell and $R$ is a lattice vector. The compact surface structure subject
of the present studies was represented by a simple hexagonal lattice. In
order to ensure that the results of the calculation accurately represent an
isolated surface, the vacuum region was varied and the convergence of the
energy was monitored. We found that a distance of five times the distance
between nearest neighbors is adequate for the present purposes. Since the
features of the computational method are extensively described in the
reference given above \cite{fhi}, we just give here a short comment on those
aspects that are relevant for the present application. This program is also
based on the density-functional formalism\cite{H&K}, where the variational
problem of a many-particle Schr\"{o}dinger equation is transformed into a
variational problem of an energy functional $E[n(r)]$, where we represent
the electron density of the system with $n(r)$, in short $n$. The energy
equation of the electronic system is then usually written as:

\[
E[n]=T^s[n]+E^H[n]+E^{e-nuc}[n]+E^{xc}[n]+E^{nuc-nuc} 
\]
where $T^s[n]$ is the functional describing the kinetic energy of a system
of non interacting electrons with density $n(r)$ , $E^H[n]$ is the Hartree
energy, calculated from the corresponding potential $V_H(\vec{r})$ and $%
E_{}^{xc}[n]$ is the so called exchange and correlation energy. In the
present case we have used the so called local density approximation (LDA) 
to the exchange and correlation energy, which applies locally the results
for the homogeneous electron gas obtained by Ceperley and Alder \cite
{Ceperley} in the parameterization of Perdew and Zunger \cite{Perdew-Zunger}%
. The remaining terms correspond to the electron-nuclei ($E^{e-n}$) and
nuclei-nuclei ($E^{nuc-nuc}$) electrostatic interaction. The electronic
density $n(r)$ is obtained through the self-consistent solution of the
corresponding Kohn-Sham equations:

\[
\left[ -\frac 12\nabla ^2+V_{ext}({\bf r})+V_H({\bf r})+V_{xc}({\bf r}%
)\right] \psi _i({\bf r})=\epsilon _i\psi _i({\bf r}) 
\]
where $V_{ext},V_H$ and $V_{xc}$ are the external potential, the Hartree and
the exchange-correlation potentials respectively, which are given by: 
\[
V_H({\bf r})=\int d{\bf r}^{\prime }\frac{n({\bf r})}{\left| {\bf r-r}%
^{\prime }\right| } 
\]
and 
\[
V_{xc}({\bf r})=\frac{\delta E_{xc}}{\delta n({\bf r})} 
\]
Thus, $n({\bf r})$ is given by: 
\[
n({\bf r})=\sum_i^{occ}f(i)\left| \psi _i({\bf r})\right| ^2 
\]
where $f(i)$ is the occupation number of state $i.\ $Instead of considering
the whole electronic system for the calculation of the properties of the
slab, the effect of the core electrons and the nuclei on the valence
electrons were replaced by suitable pseudopotentials, constructed according
to the schemes of Hamann \cite{Hamann}, which were generated and tested for
their transferability \cite{Gonze} using the code developed by Fuchs {\it et
al }. \cite{Fuchs} This involves checking for the absence of ghost states,
which may appear as consequence of using Kleinman and Bylander \cite{K&B}
fully separable {\it ab initio} potentials and monitoring the logarithmic
derivatives of the solution radial Schr\"{o}dinger equation at energies
close to that of the selected reference state.

\section{Chemical potential. Equilibrium condition. Compressibility}

When referred to the bulk deposition potential of metal $A$, the potential $%
\Phi (\Theta )$ at which a certain coverage of $A$ on a substrate $S$ is
obtained can be written as \cite{gerischer}: 
\begin{equation}
\Phi (\Theta )=\mu _{A,A}-\mu _{A,S}[\Theta ]  \label{potential}
\end{equation}
where $\mu _{A,A}$ and $\mu _{A,S}[\Theta ]$ denote the chemical potentials
of the $A$ atoms in the bulk phase and when are adsorbed on $S$
respectively. We stress the fact that this latter quantity is a function of
the coverage degree. In the usual electrochemical potentiostatic experiment,
a given potential is usually set and a coverage degree is thus obtained. The
calculations follow the reverse order. In this case a certain configuration
(coverage degree, nearest neighbor distance) of the system is considered,
and the chemical potential $\mu _{A,S}[\Theta ]$ (and thus the potential)
can be calculated.

For the discussion below, we remind the fact that we are calculating the
ground state properties of an electron gas (T=0), so that our free energies
are obtained from {\it \ energy} calculations of the system. If we denote
with $F_{A,S}$ the free energy of the substrate/adsorbate system, $\mu
_{A,S}[\Theta ]$ can be calculated from: 
\begin{equation}
\mu _{A,S}[\Theta ]=\frac{\partial F_{A,S}}{\partial N_A}  \label{mu}
\end{equation}
where $N_A$ represents the number of adsorbed atoms of type A and constant
temperature, area and volume are assumed for taking this derivative. Eq. \ref
{mu} can be also written in terms of the free energy per unit surface $%
F_{A,S}^S=F_{A,S}/S$ and the atomic surface $a$: 
\begin{equation}
\mu _{A,S}[\Theta ]=\frac{\partial F_{A,S}/S}{\partial N_A/S}=\frac{\partial
F_{A,S}^s}{\partial (1/a)}=-a^2\frac{\partial F_{A,S}^s}{\partial a}
\label{mu2}
\end{equation}
A simple relationship between the chemical potential $\mu _{A,S}[\Theta ]$
and the binding energy per atom $E_{A,S}^s$ can thus be obtained by
considering that in the present case $F_{A,S}^s=E_{A,S}^s.$ Eq. \ref{mu2}
turns into: 
\begin{equation}
\mu _{A,S}[\Theta ]=-a^2\frac{\partial E_{A,S}^s}{\partial a}  \label{mu3}
\end{equation}

This equation could also have been formulated in terms of the binding energy
per adsorbate atom $E_{bind}^{}$, which can be calculated from $E_{A,S}^s$
and the corresponding energy per unit surface of the pure substrate $E_S^s$
according to

\begin{equation}
E_{bind}^{}=a\ (E_{A,S}^s-E_S^s)  \label{ebind}
\end{equation}

In the present calculations $E_{A,S}^s$ will be considered to be equal to
the energy per unit surface of the isolated slab, since the contribution of
the substrate will be ignored.

Using now eqs. \ref{mu3} and \ref{ebind}, we obtain :

\begin{equation}
\mu _{A,S}[\Theta ]=E_{bind}-a\frac{\partial E_{bind}}{\partial a}
\label{mu4}
\end{equation}

All chemical potentials and binding energies reported in this paper have
been referred to the energy of the ion cores and the valence electrons
infinitely separated from each other. The corresponding values referred to
the energy of neutral atoms can be obtained by adding the proper ionization
energies.

We now turn to analyze the equilibrium condition in order calculate the
lattice constant of the adsorbed monolayer. The pressure of a thermodynamic
system can be calculated from: 
\begin{equation}
P=-\frac{dE}{dV}  \label{press1}
\end{equation}
where E and V are the energy and the volume of the system respectively. The
constant entropy restriction was left out because we are considering the
zero temperature case. In the case of three-dimensional solids, setting in
this equation $P=0$ provides a way of predicting the density of a metal if
an expression for E is available \cite{ashcroft}. In the present case, where
we are considering the isotropic compression of an adsorbed monolayer, it is
suitable to state the problem in terms of the area as the extensive
coordinate and its conjugate force, usually called spreading pressure \cite
{Dash}, that we denote with $\phi .$ Thus, in the zero temperature case the
analog of eq. \ref{press1} becomes:

\begin{equation}
\phi =-\frac{\partial E}{\partial S}  \label{press2}
\end{equation}
where constant temperature and volume are assumed for taking this
derivative. Thus, we shall employ the condition:

\begin{equation}
\frac{\partial E_{bind}}{\partial a}=0  \label{press3}
\end{equation}
to predict the equilibrium lattice parameter of the monolayer.

The 2D isothermal compressibility of the Tl monolayer was calculated
according to the usual definition \cite{Dash}: 
\begin{equation}
\kappa _{2D}=-\left( \frac{\partial a}{\partial \mu }\right) _T
\end{equation}
which is the same as that employed by Toney et al. \cite{Toney7} to get the
experimental values from the dependence of $a_{nn}$ on the applied potential.

\section{Results and discussion}

In addition to the logarithmic derivatives and the hardness monitoring, a
further test of the transferability of the pseudopotentials that is usually
performed consists in calculating the equilibrium lattice constant and bulk
modulus of the metal considered. We performed these calculations for Pb and
Tl, obtaining the results given in Table 1. As can be seen from the
comparison with the experimental values found in the table the agreement is
very good, not only in the lattice constant but also in the bulk modulus.

\begin{figure}
  \centering
  \subfigure[]{
   \label{fig2:subfig:a}
   \includegraphics[width=0.3\textwidth,angle=-90]{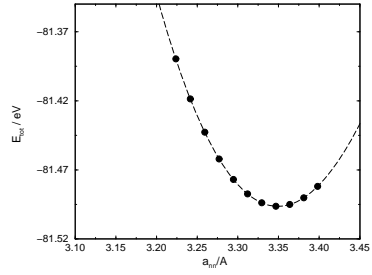}}
  \hspace{0.5in}
  \subfigure[]{
   \label{fig2:subfig:b}
  \includegraphics[width=0.3\textwidth,angle=-90]{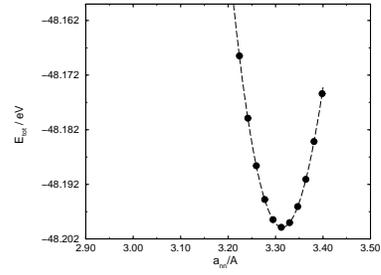}}
  \subfigure[]{
   \label{fig2:subfig:c}
  \includegraphics[width=0.3\textwidth,angle=-90]{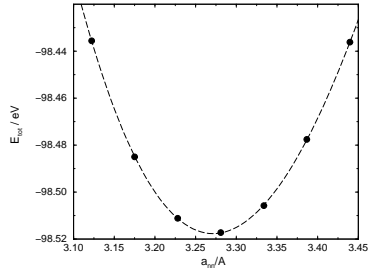}}
  \hspace{0.5in}  
  \subfigure[]{
   \label{fig2:subfig:d}
\includegraphics[width=0.3\textwidth,angle=-90]{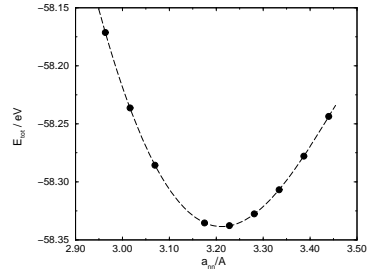}}
  \caption{Energy as a function of the distance between nearest
neighbors for single Pb and Tl monolayers isolated in vacuum. Results
obtained with local pseudopotentials, {\it 1-D} model a)Pb, b)Tl. Results
obtained with non-local pseudopotentials, {\it 3-D} model, c)Pb, d)Tl.  }
\end{figure}

Energy calculations as a function of the nearest neighbor distance $a_{nn}$
for Pb and Tl isolated slabs are shown in Figure 2. The $a_{nn}$ values at
the minimum, along with the compressibility $K_{2D}$ obtained through
numerical differentiation of these results are shown in Table 2. For the
sake of comparison, we show in both cases the results obtained by means of
the self-consistent calculation employing local pseudopotentials and the
one-dimensional solution of the effective Schr\"{o}dinger equation. The
curves obtained with the local pseudopotentials show a steeper increase of
the energy in the neighboring of the energy minimum, which is the reason for
the outcoming relatively low compressibilities. As discussed below, the
stiffness of the one-dimensional model may be understood in terms of a lack
of relaxation of the electronic density in the direction parallel to the
surface. In this respect, the introduction of the real three dimensional
structure of the electronic density brings a considerable improvement over
the calculated compressibilities.

\begin{figure}
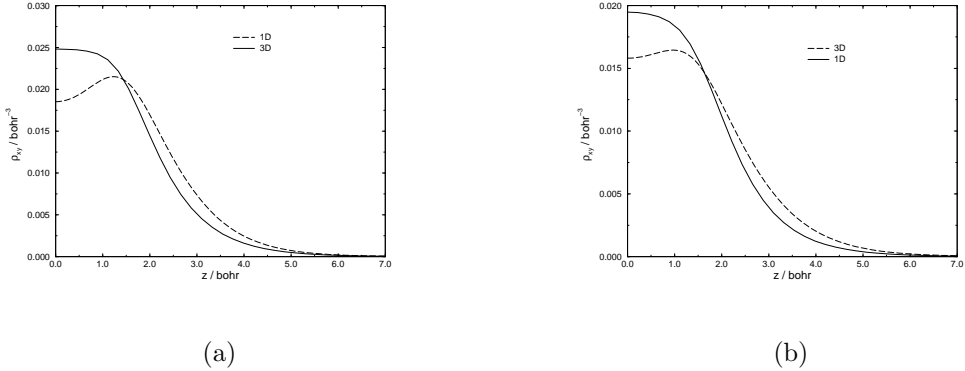

  \centering
  \subfigure[]{
   \label{fig3:subfig:a}
   \includegraphics[width=0.3\textwidth,angle=-90]{fig3a.epsi}}
  \hspace{0.5in}
  \subfigure[]{
   \label{fig3:subfig:b}
  \includegraphics[width=0.3\textwidth,angle=-90]{fig3b.epsi}}
  \caption{ Average electronic density, plotted as a function of the
distance in the direction perpendicular to the plane of the metal slab for
(a) Pb and (b) Tl. Broken line: {\it 1-D} model, full line: {\it 3-D} model.}
\end{figure}

\begin{figure}
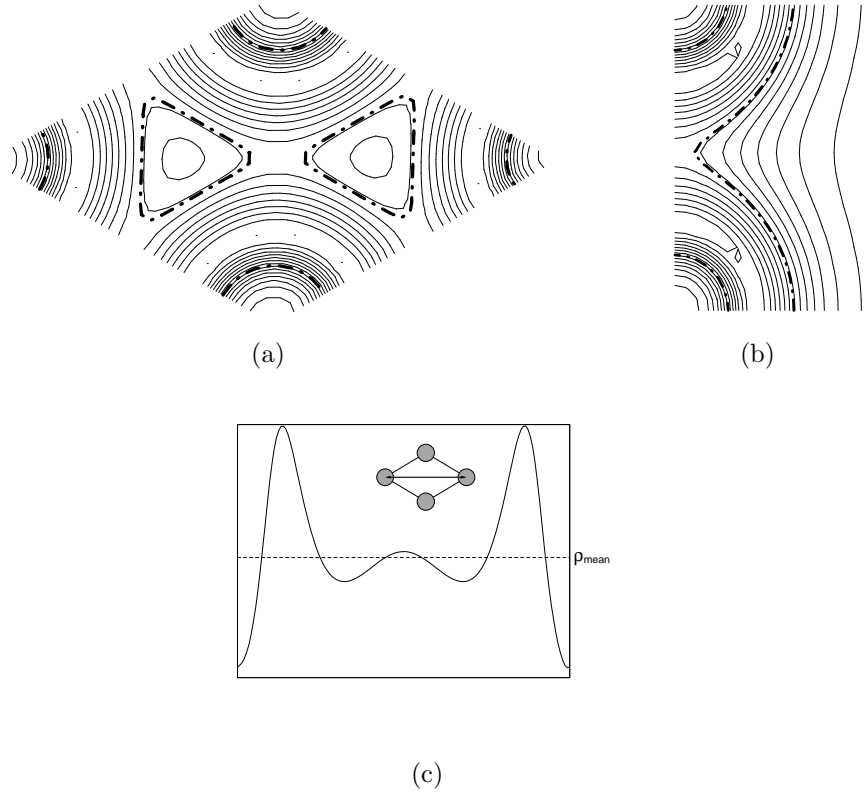

  \centering
  \subfigure[]{
   \label{fig4:subfig:a}
   \includegraphics[width=0.3\textwidth,angle=-90]{fig4a.epsi}}
  \hspace{0.5in}
  \subfigure[]{
   \label{fig4:subfig:b}
   \includegraphics[width=0.3\textwidth,angle=-90]{fig4b.epsi}}
  \subfigure[]{
   \label{fig4:subfig:c}
  \includegraphics[width=0.3\textwidth,angle=-90]{fig4c.epsi}}
  \caption{ a) Contour lines of the electronic density within a Pb
surface unit cell. b) Contour plots of the electronic density in a plane
perpendicular to the surface unit cell, containing two nearest neighbor
atoms. c) Plot of the electronic density in the line joining two atoms in
surface unit cell as shown in the inset $\rho _{mean}$ indicates the average
electronic density of bulk Pb. Thick point dashed lines in the contour plots
show the isoelectronic line corresponding to the value of the average
valence electronic density of the bulk metal. }
\end{figure}

Figure 3 illustrates some differences concerning the electronic density that
appear between the {\it 1-D model }and the more sophisticated {\it 3-D model}
calculations. In the case of the {\it 3-D} model, the electronic density was
averaged in the direction parallel to the surface (say, the x-y plane) and a
one-dimensional electronic density was obtained according to:

\begin{equation}
n(z)=\frac{\int \int n(x,y,z)dxdy}{\int \int dxdy}  \label{denav}
\end{equation}
It is observed that while the nonlocal pseudopotentials concentrate more
electronic density on the center of the metal slab, the Ashcroft -empty
core- pseudopotentials show a depletion there. On this basis only, and
thinking in terms of a free electron gas, it could appear that the Ashcroft
pseudopotentials should yield a higher compressibility, since the electron
gas with the lower electronic density should yield the higher
compressibility \cite{Ashcroft_comp}. However, the results of the
computations can be understood if we consider the electronic density in the
x-y plane, as shown in Figure 4a, where we have plotted the contour lines of
the electronic density within the layer. Figure 4b shows similar contour
plots for a plane perpendicular to the {\it x-y} plane, containing two
nearest neighbor atoms. Figure 4c shows a plot of the electronic density in
the line joining two atoms in surface unit cell as shown in the inset. It
can be appreciated that the electronic density {\it in-between }the atoms is
lower than the average electronic density $\bar{n}$, thus producing a
concomitant higher compressibility of the layer as compared with that of a
homogeneous electron gas with density $\bar{n}$.

\begin{figure}
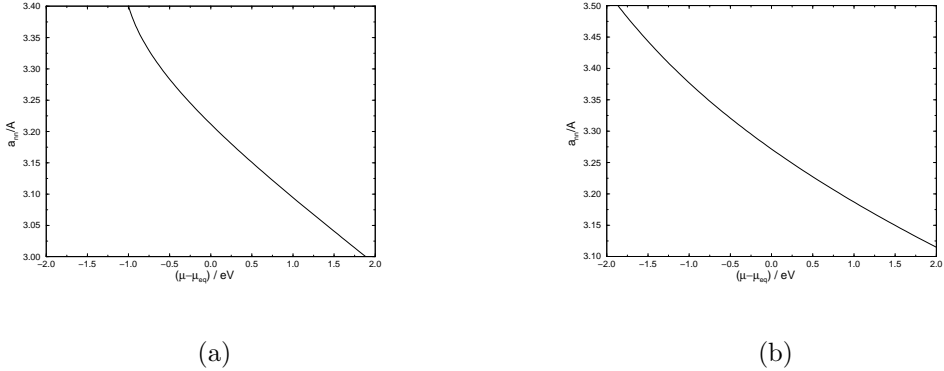

  \centering
  \subfigure[]{
   \label{fig5:subfig:a}
   \includegraphics[width=0.3\textwidth,angle=-90]{fig5a.epsi}}
  \hspace{0.5in}
  \subfigure[]{
   \label{fig5:subfig:b}
   \includegraphics[width=0.3\textwidth,angle=-90]{fig5b.epsi}}
  \caption{Nearest neighbor distance between the atoms in the
monolayer as a function of the chemical potential, a) Tl, b) Pb. }
\end{figure}

Figure 5 shows the nearest neighbor distance as a function of the chemical
potential, calculated according to the formulation of the previous section.
The chemical potentials are referred to the chemical potential of the layer
for $\phi =0$. These curves are the theoretical analog of the experimental
nearest-neighbors vs potential curves. Note that the Tl curves show a larger
curvature, a fact which is in qualitative agreement with the experimental
findings. These curves should actually not be attainable from the
electrochemical experiment for $(\mu -\mu _{eq})>0$, since this would
correspond to a negative $\phi $ (stress) of the monolayer, which would
produce an exponential increase of the number of defects and thus the
breakdown of the monolayer \cite{ashcroft}.

\section{Conclusion}

While the present theoretical compressibility values are in qualitative good
agreement with other previous estimations and show an important improvement
towards agreement with the experiment, they are still too low as compared
with the measured values. Although the presence of a substrate may deliver
some contribution to the compressibilities, we expect that an adsorbed
monolayer should be even less compressible than an isolated one, because of
the contribution of the valence electrons of the substrate. Thus, we think
that the relatively high compressibility values measured may be related to
some other aspect of the experiment so far not taken into account. A small
corrugation of the monolayer under compression could explain this fact. This
should be considered in the future modeling of these systems.

\section{Acknowledgments}

All calculations were done in a DIGITAL AlphaStation workstation donated by
Alexander von Humboltd Foundation (Germany). A fellowship (C.S.) and
financial support from the Consejo de Investigaciones Cient\'{\i}ficas y
T\'{e}cnicas de la Provincia de C\'{o}rdoba, financial support from the
Consejo Nacional de Investigaciones Cient\'{\i}ficas y T\'{e}cnicas, the
Secretar\'{\i}a de Ciencia y T\'{e}cnica de la Universidad Nacional de
C\'{o}rdoba and language assistance by Pompeya Falc\'{o}n are also
gratefully acknowledged.\newpage\

Table 1: Comparison between the theoretical predictions and the experimental
results for the lattice constant and the compressibility of the bulk metal.
The experimental results were taken from ref. \cite{Kittel}. We find a
theoretical $c/a$ ratio for Tl of 1.6; the experimental value is 1.599.
1.599.\\[0.5in]
\begin{tabular}{|l|l|l|l|l|}
\hline
Metal & a$(th)/\AA $ & a$(\exp )/\AA $ & B(th)x10$^{-12}dyn/cm^2$ & 
B(exp)/x10$^{-12}dyn/cm^2$ \\ \hline\hline
Pb & 4.98 & 4.95 & 0.530 & 0.488 \\ \hline
Tl & 3.47 & 3.46 & 0.344 & 0.359 \\ \hline
\end{tabular}

\newpage

Table 2: Nearest neighbor distance values ($a_{nn}^{\min })$ and 2D
compressibilities $K_{2D}^{\min }$ obtained at the minimum of the energy vs
nearest-neighbor curve. The theoretical values correspond to the
one-dimensional (th1) and to the three-dimensional (th2) calculation
respectively. The experimental values (exp) were taken from ref. \cite
{Toney6} and correspond to the adsorption of the metals on Ag(111) and
Au(111) surfaces.\\[0.5in]
\begin{tabular}{|l|l|l|l|}
\hline
Metal & a$_{nn}^{\min }(th1)/$\AA  & a$_{nn}^{\min }(th2)/$\AA  & a$%
_{nn}(\exp )/$\AA  \\ \hline\hline
Pb & 3.314 & 3.271 & 3.33-3.43 \\ \hline
Tl & 3.349 & 3.212 & 3.39-3.50 \\ \hline
\end{tabular}
\\[0.1in]
\begin{tabular}{|l|l|l|l|}
\hline
Metal & $\kappa _{2d}^{\min }(th1)$ & $\kappa _{2d}^{\min }(th2)$ & $\kappa
_{2d}(\exp )$ \\ \hline\hline
Pb & 0.48 & 0.53 & 1.52-1.88 \\ \hline
Tl & 0.28 & 0.72 & 1.25-1.69 \\ \hline
\end{tabular}
\\[0.1in]

\end{document}